# Raoult's Formalism in Understanding Low Temperature Growth of GaN Nanowires using Binary Precursor

*Kishore K. Madapu,[1] S. Dhara,[1],\* S. Amirthapandian,[2] S. Polaki,[1] and A. K. Tyagi[1]*

[1]Surface and Nanoscience Division, Indira Gandhi Centre for Atomic Research, Kalpakkam-603 102, India

[2]Materials Physics Division, Indira Gandhi Centre for Atomic Research, Kalpakkam-603 102, India

*Abstract*

Growth of GaN nanowires are carried out via metal initiated vapor-liquid-solid mechanism, with Au as the catalyst. In chemical vapour deposition technique, GaN nanowires are usually grown at high temperatures in the range of 900-1100 $^o$C because of low vapor pressure of Ga below 900 $^o$C. In the present study, we have grown the GaN nanowires at a temperature, as low as 700 $^o$C. Role of indium in the reduction of growth temperature is discussed in the ambit of Raoult's law. Indium is used to increase the vapor pressure of the Ga sufficiently to evaporate even at low temperature initiating the growth of GaN nanowires. In addition to the studies related to structural and vibrational properties, optical properties of the grown nanowires are also reported for detailed structural analysis.

Corresponding author's email address : dhara@igcar.gov.in

# INTRODUCTION

Since last two decades, one dimensional (1-D) semiconductor nanostructures attracted lot of attention due to their interesting physical properties and potential applications in future nanotechnology.[1] 1-D nanostructures of III-V nitride semiconductors with direct band gap, namely, InN, GaN and AlN, got special attention due to their unique optoelectronic properties. Moreover, alloys of these materials cover the band gap ranging from 0.7 to 6.0 eV showing the applications as light emitting diodes and photovoltaic cell.[2,3] Among III-V nitrides, GaN gained lot of attention due to its unique physical properties like wide band gap and high melting point for applications at elevated temperatures.[4] Moreover, GaN is considered as high power optoelectronic device material due to its high electrical breakdown field, and high carrier mobility in the high electron mobility transistor (HEMT) structures. GaN nanowire based prototype devices have already been demonstrated which include hydrogen production by water splitting,[5] field effect transistor (FET),[6] blue lasers[7] and hydrogen sensors[8] enhancing enormous interest in the research community. However, controlled synthesis in large scale and cost effective methodologies are challenges in the GaN based nano devices.

Growth of high quality GaN nanowires are reported by several techniques, namely, laser ablation,[9] metal-organic chemical vapour deposition (MOCVD),[10] molecular beam epitaxy (MBE)[11] and catalyst assisted CVD.[12] Among these techniques, CVD is a promising method for cost effective and large scale growth of nanowires.[12] In CVD technique nanowire growth is normally performed by following the well established vapor-liquid-solid (VLS) growth mechanism. Recently, nanowire growth is also reported using vapor-solid (VS) mechanism, which avoids inclusion of metal catalyst.[13] Even though Ga starts evaporating above 800 °C, growth of GaN nanowires is reported at high temperatures ranging from 900-1100 °C due to very low vapour pressure of Ga below 900 °C in CVD technique.[14] GaN is reported to sublimate at high temperatures above 800 °C,[15] and the sublimation rate increases further in case of nanostructures due to their high surface to volume ratio.[16] Moreover, decomposition temperature of GaN is 1100 °C at atmospheric pressure and under high vacuum it is 850 °C.[17] Incidentally, these decomposition temperatures are in the range of growth temperature for GaN nanowires.[9,10,12,13] Thus the reduction in the growth temperature of GaN nanowires in CVD technique carries a lot of importance. Low temperature growth at 620 °C, however was reported using Ga acetylacetonate ($Ga(acac)_3$) as the precursor in CVD technique previously.[14] Handling of the $Ga(acac)_3$, however is tricky due its poisonous and explosive nature. Moreover, reported optical properties of the grown phase were extremely poor. Recently, GaN nanowires were grown at low temperature (~650 °C) by increasing the surface energy of the Ga metal droplet with the addition of unreactive $CaF_2$ in order to increase the rate of evaporation of Ga.[18] Most recently quantum dots of GaN is also reported at 650 °C in the mesoporous silica with poor optical quality.[19]

We have approached a different methodology to decrease the growth temperature of GaN nanowires in the CVD technique using VLS mechanism with Au as catalyst. A binary alloy of Ga and In is used as precursor. We report the growth of high quality GaN nanowires even at 700 °C. Detailed structural, vibrational and optical properties have been studied. The role of In for reduction in the growth temperature of GaN nanowires is discussed in the frame work of Raoult's law.

**EXPERIMENTAL**

Growth of nanowires was carried out by customized CVD set up with a horizontal tube furnace.[12,13] Pure Ga (99.999% Sigma Aldrich) and In (99.999% Sigma Aldrich) metals were used as the source material in the mass ratio of 1:3. Before mixing the two metals, In shots were cleaned with 10% HCl solution for etching the native $In_2O_3$ layer to ensure the oxygen free In surface. Polished 10 × 10 mm² dimension Si(100) wafers were used as substrate. Cleaned Si substrate was coated with Au (thickness ~ 3 nm) in a thermal evaporation chamber with a base pressure of $10^{-6}$ mbar. The thickness was monitored by quartz crystal microbalance. Source material and Au coated substrate was placed inside the ceramic boat with separation of 2 cm from the source in downstream. Ceramic boat with source and substrate was transferred to 1 inch quartz tube and later tube was degassed by rotary pumping. Growth of nanowires was carried out at three different temperatures of 700 °C, 750 °C and 800 °C for 7 hrs each in the atmospheric pressure. Temperature of the furnace was increased to growth temperatures at a rate of 20 °C per minute. Reactive ultra high pure $NH_3$ (99.999%) was introduced at growth temperatures with a constant flow rate of 100 sccm. High pure Ar was purged with a flow rate of 100 sccm until the growth temperature was reached. After the reaction, furnace was cooled to room temperature under the $NH_3$ atmosphere.

Field-emission scanning electron microscope (FESEM; Zeiss, SUPRA 55) was used to study the morphology of the nanowires. Structural analysis was performed using high resolution transmission electron microscopy (HRTEM; Libra 200 Zeiss), selected area electron diffraction (SAED) and X-ray diffraction (Brucker, D8 Discover) techniques. Compositional studies were also studied using energy dispersive x-ray spectroscopy (EDS). Vibrational properties were investigated by using micro-Raman spectrometer (inVia; Renishaw, UK) in the back scattering configuration with excitation of 514.5 nm $Ar^+$ laser, 1800 gr/mm grating for monochromatization and thermoelectric cooled CCD as detector. The optical properties of the samples were studied using temperature dependent photoluminescence (PL) spectroscopy with excitation of 325 nm He-Cd laser, and 2400 gr/mm grating.

**RESULTS AND DISCUSSION**

Corrugated morphology of nanowires grown at three different temperatures is shown in Figure 1. It is observed that diameter of the nanowires is in the range of 50-75 nm and length is up to few micro meters. The corrugation in these nanowires can be due to either inhomogeneous nucleation or enhanced polycrystallinity at low growth temperature,[20] invoking faceted morphology. Compositional study of nanowires was typically studied by EDS along with the morphological analysis. EDS analysis reveals that Ga and N exist along with the trace amount of oxygen (supporting information Figures S1). Ga to N atomic percent ratio is found to ~ 0.9 for these samples showing reasonable stoichiometry in the GaN phase, with the consideration of various factors of limited capability of absolute quantification in the EDS technique and in the presence other impurities, primarily as adsorbed species. Presence of oxygen with atomic percent ratio of O to Ga ~ 0.15, as estimated using EDS spectra, may originate either from the oxidized Si substrate at high temperatures or as absorbed specie on the nanowire surface. However, we have not found any trace amount of In incorporation in the material showing that In is not actively participating in the growth of the nanowires.

Typical low magnification TEM morphology of nanowires grown at 750 $^{o}$C is shown (Figure 2a) with similar corrugated features as observed in the FESEM (Figures 1a-c) studies. Spotted SAED pattern of the nanowires are shown (Figure 2b) with zone axes along [01-10] corresponding to the wurtzite GaN. Absence of ring pattern form these corrugated GaN nanowires proves presence of textured grains. XRD study also show presence of wurtzite GaN phase in these nanowires for a sample typically grown at an optimum substrate temperature of 750 $^{o}$C (supporting information Figure S2). HRTEM image of the nanowire (Figure 2c), shows lattice planes parallel to the nanowire growth axis. Spacing between planes are ~ 0.276 nm which belongs to (10-10) of GaN. So the growth direction is expected in the direction of [-12-10] which is perpendicular direction of [10-10], as inscribed in Figures 2c and 2d. Detail HRTEM analyses show (Figures 2c and 2d) presence of multiple grains in these nanowires. These grains, however are having the same orientation of [-12-10].

Vibrational properties of nanowires are studied using Raman spectroscopy with 514.5 nm excitation. Raman spectra (Figure 3) for these nanowires grown in the temperature range of 700 - 800 $^o$C show symmetry allowed modes ~ 565 and 721 cm$^{-1}$ corresponding to $E_2$(high) mode and longitudinal-optical mode of $A_1$ symmetry [$A_1$(LO)] of wurtzite GaN, respectively.[21] Peak centred at 521 cm$^{-1}$ corresponds to Si. Raman spectra of GaN nanowires grown at 800 $^o$C show the distinguished sharp Raman peaks corresponding to wurtzite GaN. In case of nanowires grown at 700 $^o$C Raman spectrum is broadened, which may be due to low crystallinity in the grown phase at low temperature. Absence of $E_2$(high) mode may also be noted in this regard.

Low temperature (80K) PL study was used for analysing optical properties of nanowires to shed light on the possible mechanism of low temperature growth of GaN in presence of In. Nanowires grown in the temperature range of 700-800 $^o$C shows the strong emission in the energy range of 2.65 to 3.6 eV (Figure 4). The peaks are fitted only for Gaussian shape describing luminescence phenomenon, leaving the sharp peaks which are mixed with Lorentzian shape as physical origins are not the same for these peaks and will be evident in the following discussion. In case of nanowires grown at 800 $^o$C, two prominent emission peaks at 3.47 eV and 3.38 eV are observed (Figure 4a). The symmetric peak at 3.47 eV with a typical width of 61 meV may correspond to the free exciton (FE).[22] FE emission line was followed by a broad peak at ~3.38 which can be attributed to the free exciton-phonon [$A_1$(LO)~ 90 meV] replica (FE-LO).[22] The weak and broad peak at 2.9 eV may be assigned as blue luminescence (BL) band originating from the transitions of carriers in the shallow donor or conduction band to deep acceptor band.[23] A shallow donor and deep acceptor (~0.4 eV above acceptor level)[24] states may arise due to presence of N antisites ($N_{Ga}$). However, role of deep acceptor state (~0.3 eV above acceptor level)[23] owing to Ga vacancies ($V_{Ga}$) cannot be ruled out in these samples grown at low temperatures where Ga vapour pressure is expected to be low.[14] Similar features with broadening of the peaks with diminished BL peak intensity are observed for the samples grown at 750 $^oC$ (Figure 4b). As reported earlier, we also found that BL band intensity was temperature independent in the range of 100-200K.[23] At higher temperatures thermal quenching above 200K was observed in BL intensity due to the redistribution of holes which could escape to valence band from the acceptor states.[23] The BL band is observed to be quenched for samples grown below 800 $^oC$, as formation of $N_{Ga}$ may not be favoured for the sample grown below 800 $^oC$. Moreover, PL intensity is found to be strong in case of 800 $^oC$ grown sample as crystalline GaN phase is expected at high growth temperature. Low temperature (80K) PL spectrum of nanowires grown at 700 $^oC$ show broad emission around 3.29 eV and is found to be stronger than FE emission line at 3.47 eV (Figure 4c). Broad emission around 3.29 eV can be attributed to free-to-bound (FB) emission line which is mediated by the free to a deep acceptor state.[13] The strong FB intensity for sample grown at 700 $^oC$ may be due to the formation of deep acceptor state of $V_{Ga}$,[25] as N-rich condition prevail in the availability of diminished Ga vapour with decreasing growth temperature.[14] Deep acceptor state is energetically favoured to be formed in presence of $V_{Ga}$ which is responsible for the observed FB transition. Resonance Raman modes (best described as Gaussian mixed with Lorentzian shape) up to 3rd order corresponding to $A_1$(LO) phonon were observed along with PL spectra of samples grown below 800 $^oC$. Excitation above the band gap mediate coupling of LO phonon with the electron in the conduction band due to Frolich interaction, invoking non-zone centre higher order phonon modes in the Raman spectra.[25] Low PL intensities for the samples grown at low temperatures make the resonant modes visible in Figures 4b and 4c.

The band gap of wurtzite phase of pure bulk GaN is 3.43 eV at room temperature and the value is not expected to vary with mere reduction of dimension unless the size in one of dimensions is reduced below the excitonic Bohr radius of GaN ~11 nm.[26] Figure 5 shows temperature dependent PL data of GaN nanowires grown at 800 °C. At room temperature, a considerably broad peak at 3.39 eV dominates, which may correspond to the near band edge emission (NBE).[27] The peak gets blue shifted to 3.47 eV, corresponding to FE emission line with lowering of temperature. The FE emission line is observed to be suppressed as the temperature increases. This may be due to the delocalization of carriers at defect bands where exciton binding is not efficient. The FE-LO peak at 3.38 eV quenches above 80K. The broad NBE band around 3.3 eV and BL band at ~3 eV evolve with increasing temperature, clearly showing involvement of defect related states in the electronic transition of carriers. However, we have not found any effect of indium in the PL studies in the absence of tuning of NBE peak position. Considering the energetic of electronic exchange processes involved, a schematic band diagram is shown (Figure 6) depicting the possible transitions as discussed above. These include emission characteristics of FE ~ 3.47 eV, NBE ~ 3.39 eV, FB band ~ 3.3 eV, and BL band ~2.9 eV with a shallow donor level and deep acceptor state at 0.4 eV above acceptor level corresponding to $N_{Ga}$ native defect. A deep acceptor state at 0.3 eV above acceptor level corresponding to $V_{Ga}$ native defect is also depicted.

Source, precursor and metal catalyst play crucial roles in the growth of nanowires in CVD technique, especially in case of VLS growth mechanism. A sufficient amount of precursor material is necessary in the vapor phase to initiate the growth of nanowires at any particular temperature. Amount of precursor in vapor phase depends on the vapor pressure of that material at a given temperature. Usually growth of GaN is carried out at high temperature so that sufficient amount of precursors is available in the vapour phase.

However, growth at high temperatures is having some disadvantages towards the final product formation. In case of GaN, we have already discussed the disadvantages of high temperature growth with possible disintegration of the compound. Vapor pressure of Ga is of the order of $10^{-3}$ Torr at temperatures above 900 °C and this is sufficient to initiate the growth of nanowires. One can achieve, however the growth at temperatures lower than 900 °C if the effective vapor pressure of Ga is increased.[18] We have achieved the growth of the nanowires at low temperatures as low as 700 °C using binary alloy of Ga-In as the precursor material.

Growth of GaN nanowires at low temperature may be proposed in the frame work of Raoult's law which is applicable only for ideal solutions. Figure 7 shows the linear relationship of the vapor pressure of the ideal solution with mole fraction of components, schematically.

$$p_1 = x_1 p_1^* \qquad (1)$$

$$p_2 = x_2 p_2^* \qquad (2)$$

$$p_{total} = p_1 + p_2 \qquad (3)$$

Equations (1)-(3) express the vapor pressure relations of the solution where $p_1$ and $p_2$ are the vapor pressure of 1st and 2nd components in a solution with mole fractions are $x_1$ and $x_2$, respectively. While $p_1^*$ and $p_2^*$ are vapor pressure of the 1st and 2nd components, respectively, in their pure form; $p_{total}$ is the total vapor pressure over the solution. Overall vapor pressure of the solution depends on the mole fraction of the individual components. Total vapor pressure of the solution is given by the equation (3). In case of the real systems Raoult's law shows two types, namely, positive and negative deviations. In case of positive deviation from Raoult's law, solution shows the vapor pressure higher than that predicted by the equation (3) and vice versa for negative deviation (schematic in Figure 7). Ga-In system forms a eutectic alloy at ~ 15 °C.[28] We can treat the solution of Ga-In as an ideal binary solution as the growth temperature of nanowires is far away from the eutectic region of the Ga-In system. Indium possesses higher vapor pressure than that of Ga at a given temperature. Applying the Raoult's law for Ga-In system, solution of Ga-In possesses vapor pressure more than Ga (Figure 7). At a given temperature In lifts the vapor pressure of the Ga so that the effective vapor pressure Ga is high enough to initiate the growth. Lowering of GaN growth temperature, as it is claimed throughout the present study, is possible due to the positive deviation in the Raoult's law (Figure 7). This deviation further increases the effective vapor pressure of Ga even at low temperatures. Positive deviation from Raoult's law is possible if inter-atomic interactions between In-Ga are weaker than the interactions between In-In and Ga-Ga atoms. Svirbely and Read[29] calculated the activity curves of Ga-In system indirectly from electromotive force measurements of Ga-In-Zn, In-Zn, and Ga-Zn systems. Calculated activity curves show that Ga has a strong positive deviation from Raoult's law over the entire compositions range and In has a strong positive deviation from Raoult's law in the composition range of $0 < X_{In} < 0.50$. Later it was confirmed experimentally by Macur *et. al.*[30] Activities of Ga and In in Ga-In system are derived from effusion measurements utilizing the multiple Knudsen-cell effusion technique. The study reported that positive deviation from the Raoult's law occurs for both the components. We have tried to grow nanowires in the same temperature range using Ga metal in the absence of In to ensure the role of In for the reduction in the growth temperature. We have not found any yield of nanowires even at 800 °C with same growth conditions.

**CONCLUSION**

In conclusion, we have grown the GaN nanowires at low temperatures, as low as 700 °C in the CVD technique. The growth temperature is reduced by almost 200 °C using binary alloy of Ga-In as the precursor. TEM analysis reveals that these low temperature grown nanowires are crystalline in nature having textured grains. High electronic quality of the GaN nanowires is confirmed from the temperature dependent photoluminescence studies. Presence of In is not found in the optical studies. As the growth temperature decreases, formation of Ga vacancies increase in the N-rich conditions showing characteristic features of free-to-bound and blue band emissions. Growth of the GaN nanowires at low temperature is explained in the ambit of Raoult's law with the fact that In has higher vapour pressure than that of Ga. Lowering the growth temperature of the GaN nanowires further enhances the interest for GaN device application.

## ACKNOWLEDGEMENTS

We would like to thank T. R. Ravindran of Condensed Matter Physics Division, IGCAR for his help in Raman studies. We also thank Ravi Chinnappan of Metal Physics Division, IGCAR and Arindam Das of SND, IGCAR for their valuable suggestions during the preparation of manuscript.

## SUPPORTING INFORMATION

EDS and XRD studies for compositional and structural analyses of GaN nanowires, respectively. This information is available free of charge via the Internet at http://pubs.acs.org

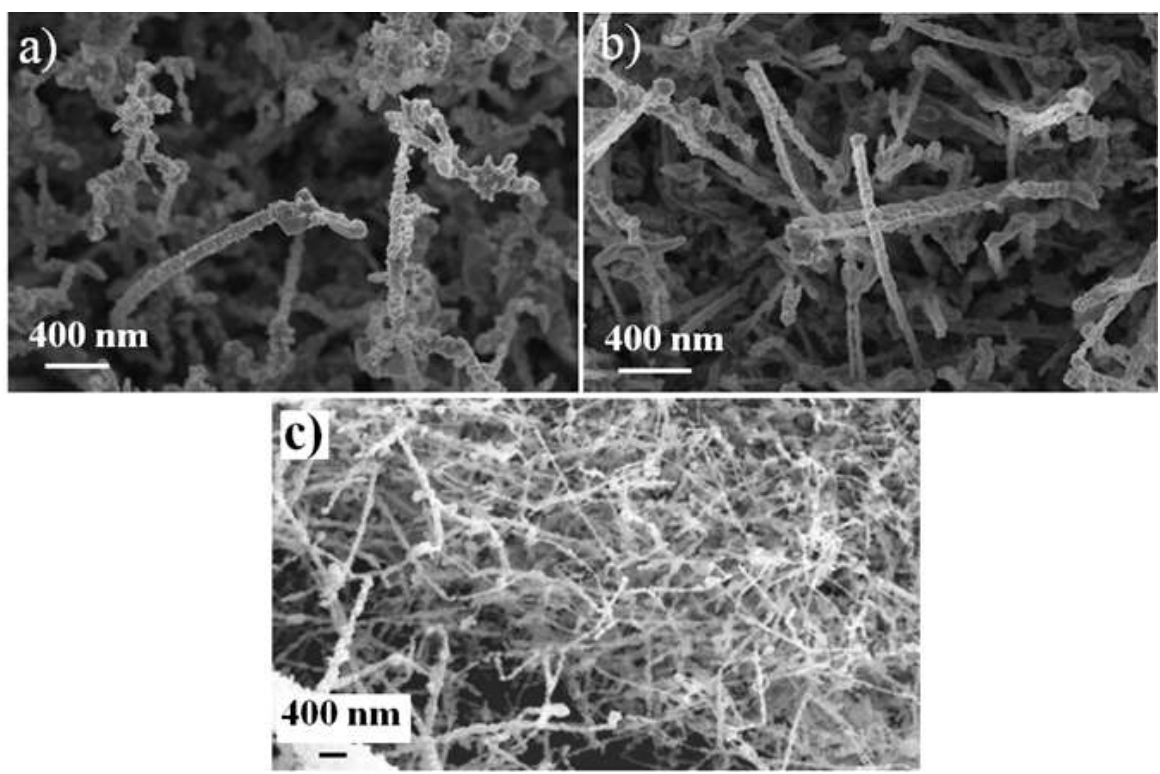

**Figure 1.** FESEM images of the GaN nanowires grown at low temperatures of a) 700 $^{o}$C b) 750 $^{o}$C and c) 800 $^{o}$C with corrugated morphologies.

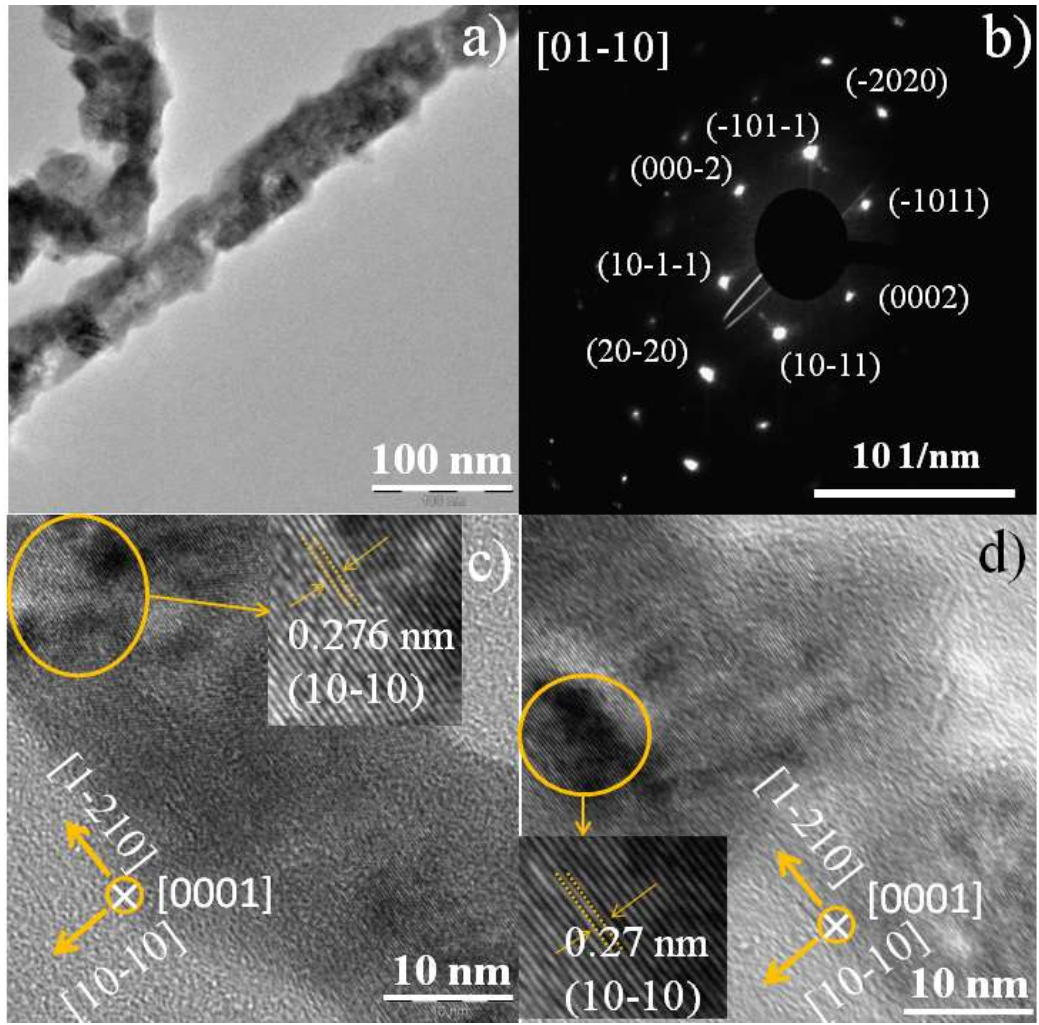

**Figure 2.** a) Typical low magnification TEM image of the GaN nanowires grown at 750 °C. b) Selected area electron diffraction pattern taken along the [01-10] zone axes. c) and d) HRTEM images of the GaN nanowire with detail lattice spacing and growth direction inscribed in the left side.

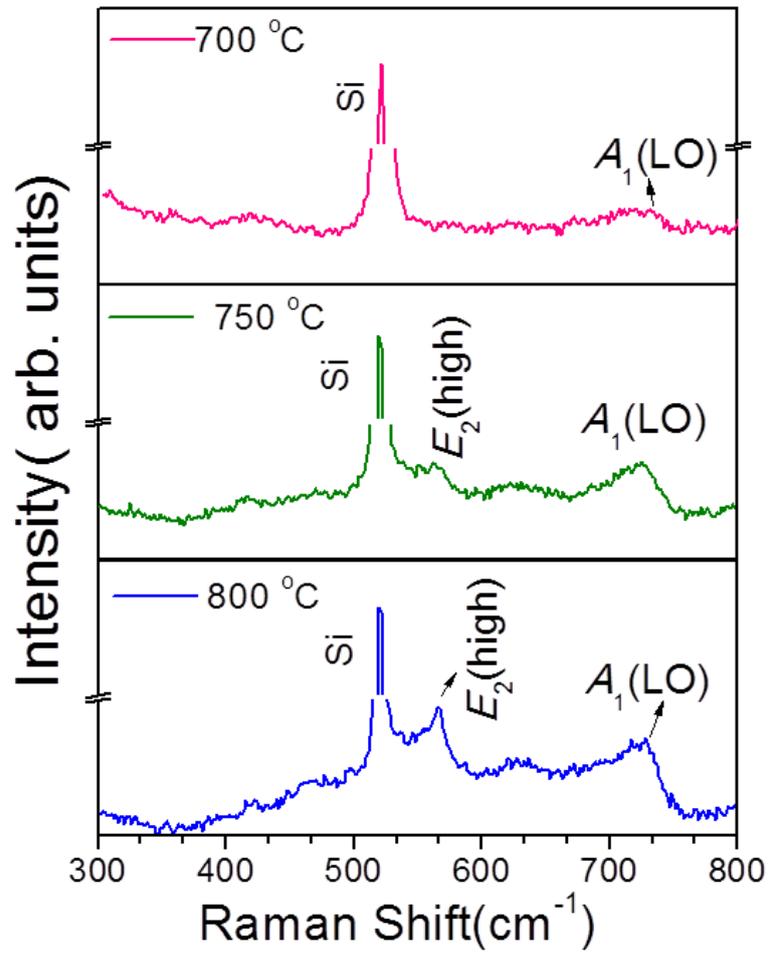

**Figure 3**. Raman spectra of GaN nanowires grown at three different temperatures of 700 °C, 750 °C and 800 °C showing symmetry allowed Raman modes.

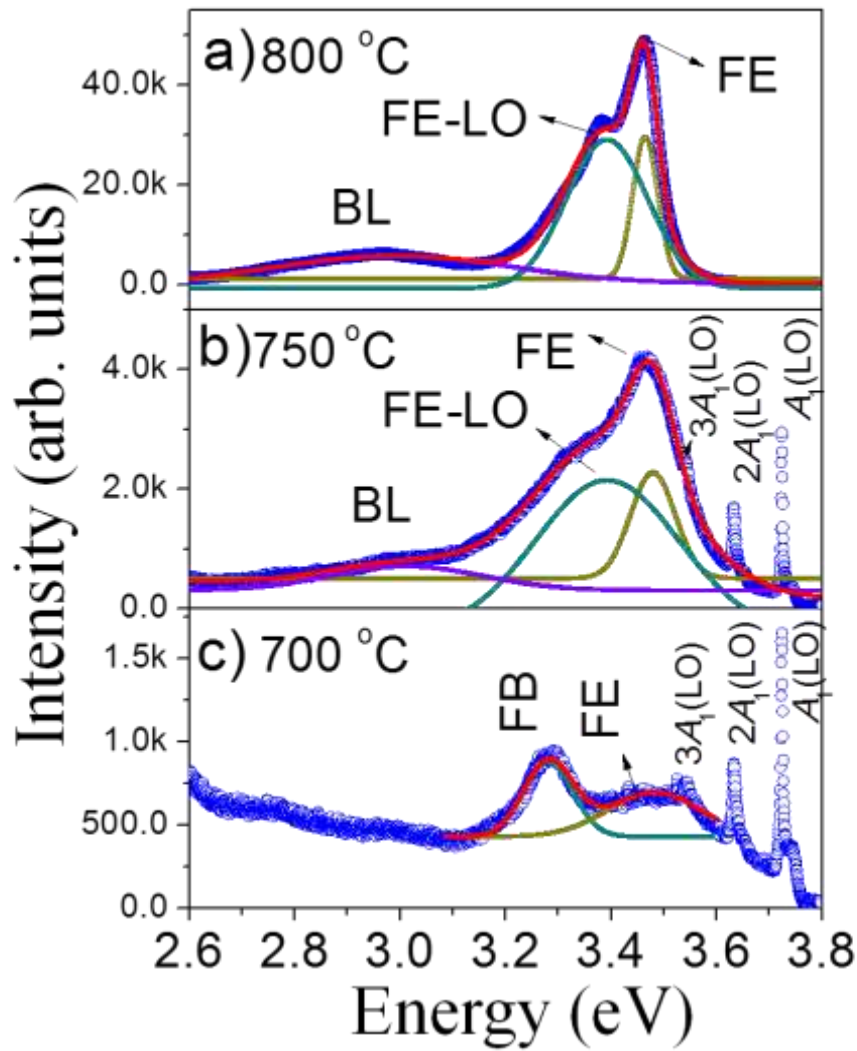

**Figure 4**. Photoluminescence spectra of GaN nanowires measured at 80K and grown at three different temperatures of a) 800 °C, showing a strong emission of free exciton (FE) and followed by its phonon replica (FE-LO) with blue luminescence (BL) band observed with considerable intensity; b) 750 °C, similar features as observed for 800 °C grown samples except with a diminished BL band intensity and c) 700 °C, showing free-to-bound (FB) transition relatively stronger than FE. The peaks are fitted for Gaussian shape alone to identify the luminescence features.

19 | P a g e

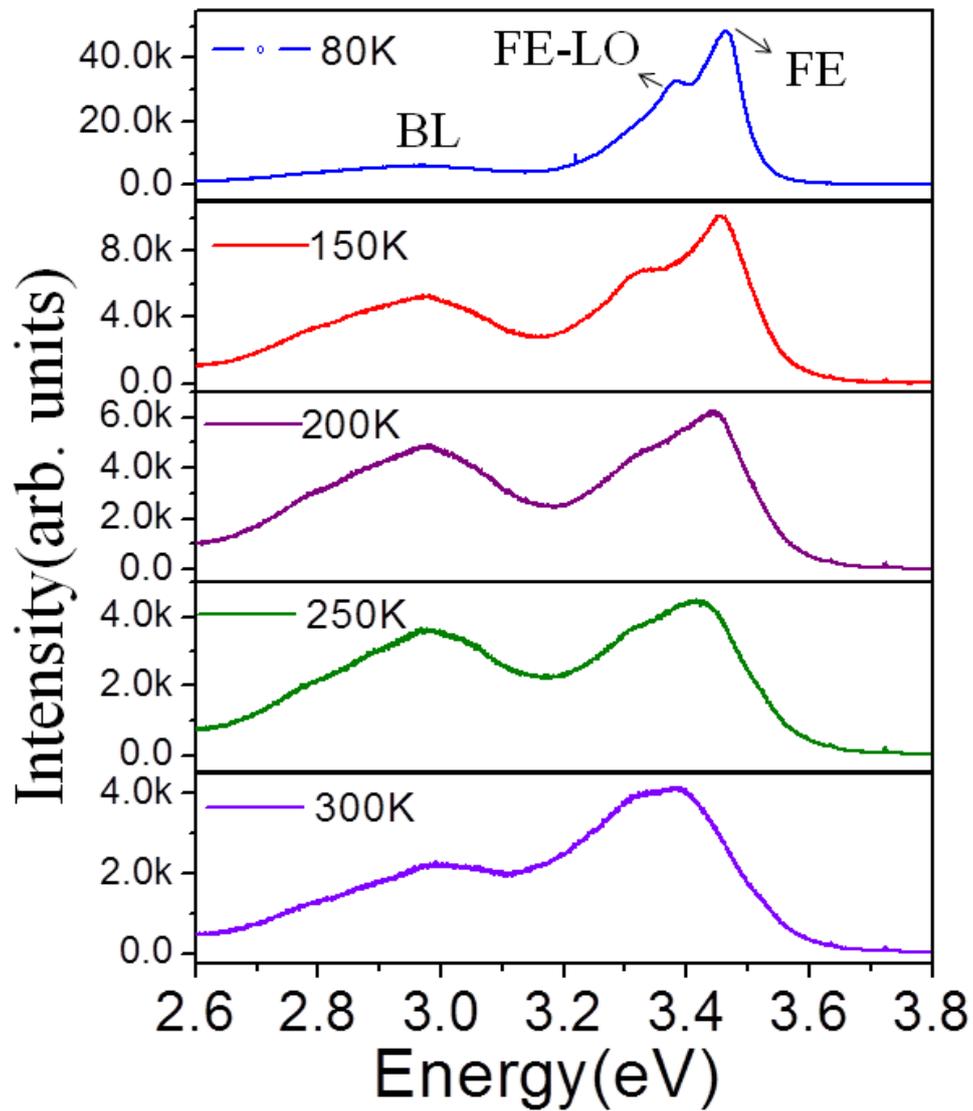

**Figure 5**. Temperature dependent photoluminescence spectra of the GaN nanowires grown typically at 800 °C. Blue shift was observed in the near band edge (NBE) as the temperature decreases to low temperatures.

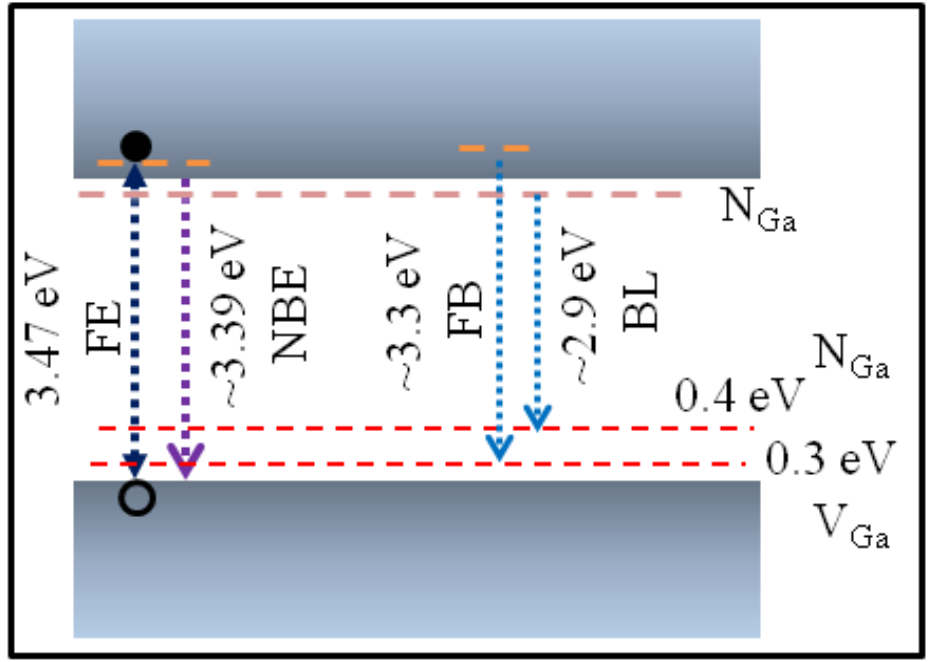

**Figure 6**. Schematic energy level band diagram corresponding to different electronic transitions of free exciton (FE), near band edge (NBE), free-to-bound (FB) and blue luminescence (BL) bands observed in GaN nanowire samples in the present study. A shallow donor level and deep acceptor state at 0.4 eV above acceptor level corresponding to N anitisite ($N_{Ga}$) along with deep acceptor state at 0.3 eV above acceptor level corresponding to Ga vacancy ($V_{Ga}$) native defects are also depicted.

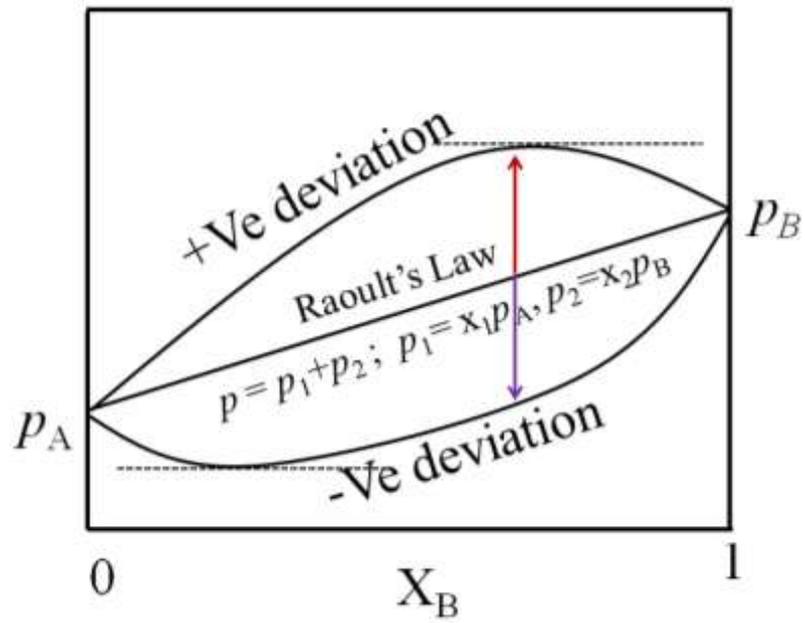

**Figure 7**. Schematic diagram of Raoult's law and its possible deviations invloving Ga-In alloy.

**Supporting information :**

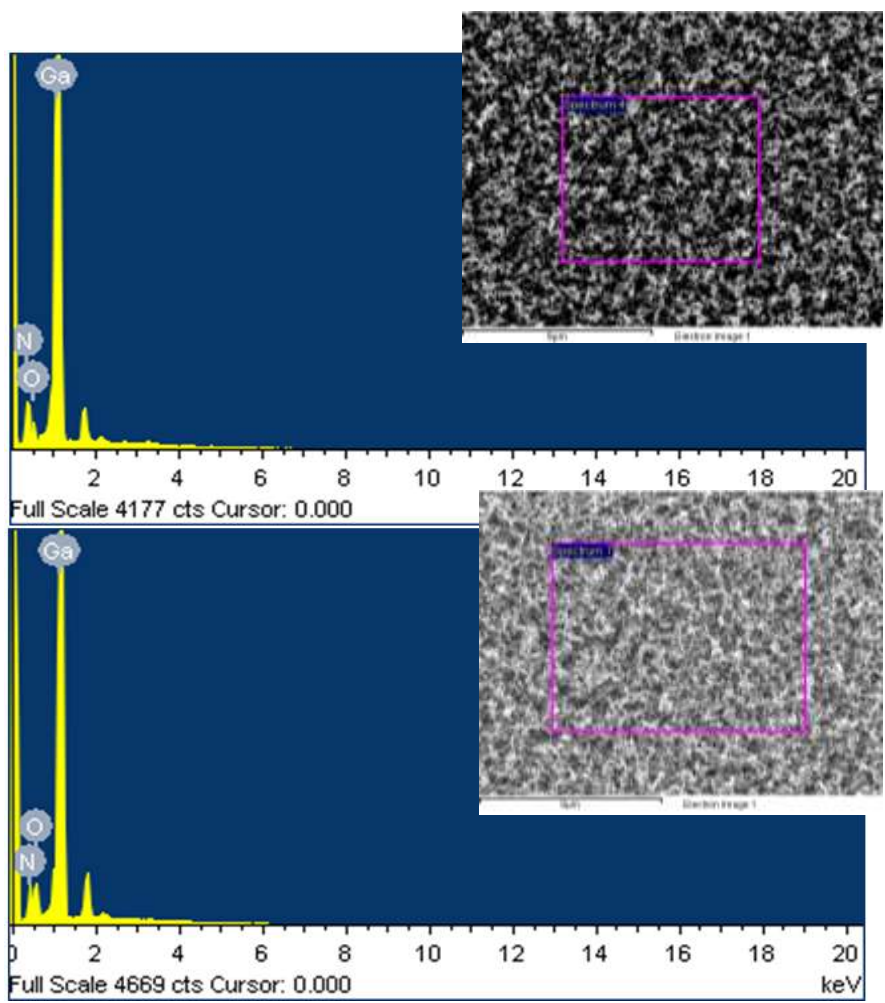

Fig. S1. Energy dispersive spectroscopic analysis of GaN nanowires grown at 800 °C and 750 °C showing Ga and N only along with O originating either from oxidized Si substrate or adsorbed on the nanostructures. Presence of Si ($K_\alpha$) at 1.74 eV and a trace amount of catalytic Au (*M*) at 2.12 eV are also observed. A tiny feature at ~ 3.4 eV in the EDS spectra is much beyond the In ($L_\alpha$) position of 3.286 and the peak intensity is too small to be identified as In. The SEM image and the region over which the data is collected, marked as square, are also shown in the insets of the respective EDS spectrum.

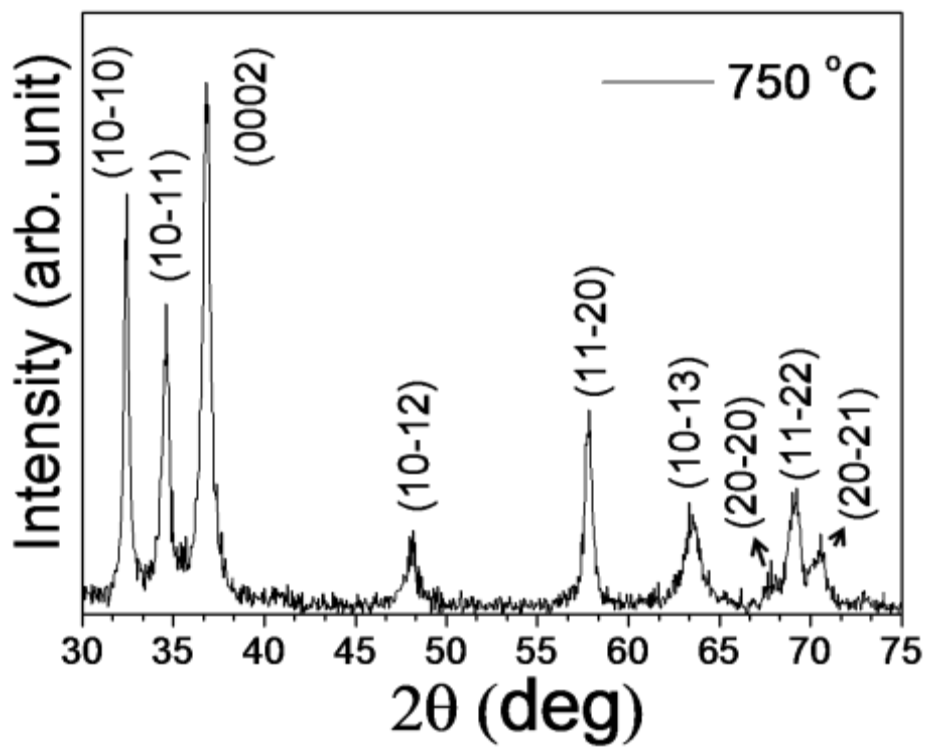

Fig. S2. XRD a sample typically grown at an optimum substrate temperature of 750 °C show presence of wurtzite GaN phase in these nanowires. While other planes are also present, predominant crystalline planes correspond to (10-10), (10-11) and (0002) of wurtzite GaN phase (JCPDS # 02-1078).